\documentclass[preprint,prd, nofootinbib]{revtex4-1}
\usepackage{color}
\usepackage{graphicx}
\usepackage{epsfig}
\usepackage{subfig}
\usepackage{lipsum}
\usepackage{ctable}
\usepackage{hyperref}
\usepackage{physics}
\usepackage{graphicx}
\usepackage{dcolumn}
\usepackage{bm}

\begin{document}

	\title{Wormholes in the Non-minimally  Coupled  Gravity with Electromagnetism}

	\author{ \"{O}zcan SERT}
	\email{osert@pau.edu.tr}
	\affiliation{Department of Physics, Faculty of  Sciences, Pamukkale	University,  20070   Denizli,
       T\"{u}rkiye  }
	

	\date{\today}

	\begin{abstract}

\noindent

Since the general relativistic approach requires exotic matter with negative energy density,
constructing wormholes containing realistic matter is a crucial challenge.  Therefore, extending General Relativity to non-minimal cases may be an alternative option. In this paper, we   investigate   wormholes which are supported by  the non-minimally coupled electromagnetic fields to gravity in $Y(R)F^2$ form where $F^2$ is the Maxwell invariant and $Y(R)$ is a function of the Ricci scalar $R$.
We consider both electrically and magnetically charged cases. Then we give some exact asymptotically AdS/dS and flat solutions for traversable wormhole geometries. By discussing whether these obtained geometries satisfy the energy conditions and vanishing sound speed condition,  we determine  stable solutions containing realistic matter   for some parameter values.

	\end{abstract}
	
	\pacs{Valid PACS appear here}
	\maketitle

\section{Introduction}
 
 Wormholes  are one of  the most fascinating  prediction of General Relativity (GR). Inspired by the recent observations of the other predictions of GR such as black holes and gravitational waves,   the possibility of their existence  encourages  scientists to  study them increasingly.   
They are  tunnels   which can  connect two  faraway  regions of the same universe or   two different universes with a shortcut. They might even turn into time machines that could allow time travel for highly advanced civilizations \cite{MorrisThorne1988,MorrisThorneYurtsever1988}.

 Wormhole structures   were firstly noticed by Ludwig Flamm in 1916  as a bridge between black hole and hypothetical white hole \cite{Flamm1916}   
  and rediscovered by Einstein and Rosen \cite{ER1935} in 1935 looking for a solution to the two-body problem in General Relativity and  discussing whether material particles are singularities in the field.  In order to find a unified description   of gravity and electromagnetism,   Einstein and Rosen 
 tried  to obtain a non-singular description of material particles.
 They obtained regular  spherically symmetric and  static solutions by modifying General Relativistic field equations (by multiplying them the square of the determinant of the  metric) and  changing variable in the  solutions.  
 Thus, they had  a solution to this singularity problem by replacing  an elementary  particle with a finite bridge. They thought that the elementary particle corresponded to the bridge connecting the geometries (Schwarzschild or Reissner-Nordström)  at   two opposite  ends  or sheets, known as Einstein-Rosen (ER) bridge. 
Recently, after many years,   the ER bridge   was associated with  Einstein-Podolsky-Rosen paradox  as "ER=EPR" conjecture   by  Maldacena and Susskind \cite{Maldacena2013}  in order to explain the connection between  any two entangled quantum systems. According to this conjecture the  
 ER bridge  (or wormhole geometry connected by two quantum systems) \cite{ER1935} is equivalent to the  EPR entanglement  of two quantum systems \cite{EPR1935} and this led to the idea that quantum particles could communicate via wormholes recently.

  This Einstein-Rosen bridge was  named as  wormhole by Wheeler for the first time   in the 1950s \cite{Wheeler19572,Wheeler19571}. 
However this is not a traversable  wormhole as it does not allow  two-way travel due to the presence of the event horizon \cite{Fuller1962}.
 The traversable wormholes allowing travel among distant  universes first constructed  by Morris and Thorne \cite{MorrisThorne1988}  which are free from singularities and event horizons. This work further developed  and revealed that the traversable wormhole could allow time travel by Morris, Thorne, and Yurtsever \cite{MorrisThorneYurtsever1988}. Later, many general relativistic wormhole studies \cite{Hawking1988,Visser1989, Visser2002,Visser2021,Curiel2017,Hochberg1997,Lemos2003,Damour2007,Lobo2005,Sushkov2005,Bueno2018,JavedOVgun2022,Ovgun2016,RahamanPant2019,Garcia2012,Lobo2008,Nashed2023} inspired  by these seminal papers. However, these traversable wormhole solutions can be constructed by exotic matter with negative energy density or pressure which violate the null energy condition in General Relativity.

 The null energy condition states that $\tau_{ab}k^ak^b \geq 0 $ where $\tau_{ab}$ matter energy-momentum tensor   and  $k^a$  any null vector.  The weak energy condition states that $\tau_{ab}u^au^b \geq 0 $  for any time-like vector  $u^a$, see e.g. a recent review \cite{Kontou2020}  for more details on energy conditions.
 The first condition corresponds to  $\rho + p \geq 0$, while the second  to  $\rho + p \geq 0$ and   $\rho\geq 0 $ for an isotropic fluid with energy density  $\rho$ and  pressure $p$.
 It is  a challenging problem to find  traversable solutions without negative energy density and satisfying the null and weak energy conditions.  In order to construct  traversable wormhole solutions with a realistic matter,   alternative modified theories of gravitation can be considered. Some  wormhole solutions  satisfying the energy conditions  can be found 
for the non-minimal
curvature-matter coupled theories \cite{Montelongo2011,Garcia2010},
Einstein Gauss-Bonnet theory \cite{Bhawal1992,Mehdizadeh2015}, Brans-Dicke theory \cite{Anchordoqui1997},
$f(R)$ \cite{Lobo2009,Harko2013,Falco2021} and $f(R,T)$ \cite{Rosa2022} modified theories of gravity. Furthermore,
some various wormhole solutions   were given and investigated in beyond the Riemann geometry such as $f(T)$-gravity where $T$ torsion invariant, $f(Q)$-gravity where $Q$ nonmetricity invariant and more general metric-affine theories in
\cite{Sebastiani2023,Sebastiani2024,Falco20212,Falco2023,Agrawal2024,Nashed2009}, see also the
extensive catalog \cite{Stephani2003tm}.


A natural generalization of the Morris-Thorne wormhole is to study charged cases that arise in the presence of electric and magnetic fields. In particular, the presence of intense electrical charge in the throat of wormhole creates a repulsive force, preventing wormhole from collapsing and helping its throat to open. Moreover, charges support  stability  of  wormhole playing the role of exotic matter \cite{Kim2001,Kuhfittig2021,Ernesto2004}.  
Furthermore,  charged  modified gravity models  induce  a surrogate to the exotic matter by increasing  energy of wormholes, therefore they support the stability of wormholes \cite{Ovgun2023,Sharif2023,Jusufi2017,Moraes2019,Sharif2021}.



There are important clues that minimal coupling is not sufficient in classical or even quantum  field theory,  which  in general there is no clear classification of higher-dimensional interactions into tree and loop operators based on the minimal coupling \cite{Jenkins2013}.  
It is very important to examine the non-minimal coupling of fundamental fields such as scalar, vector and spinor fields to gravity.
A well-known example of these non-minimal couplings is the $R\phi^2$  or more generally $RV(\phi)$ type coupling, where the scalar field  $\phi$ is non-minimally coupled to gravity encoded by the curvature scalar  $R$.
This type coupling  is  necessary for renormalization of the scalar field in curved space\cite{Birrell}.
Also such couplings can be used to describe inflation and  the early expansion of the universe,  for reviews see e.g. \cite{Lyth1999,Bassett2006}.

In order to describe astrophysical events in extreme situations where gravitational and electromagnetic fields are very intense  such as neutron stars and charged black holes, deviations from the classical laws can be considered that would allow new types of interactions between electromagnetism and gravity.
This kind of interactions in $RF^2$ form was introduced by Prasanna \cite{Prasanna1971} firstly to obtain null electromagnetic field in conformally flat   spacetime,  where $F^2$ is Maxwell invariant.   
 Later such  $RF^2$-type coupled invariants in four dimensions only lead to second order field equations  were determined and classified by Horndeski \cite{Horndeski1976}. 
Furthermore in quantum field theory, they can also occur at the 1-loop level of the photon vacuum polarization in the curved space background 
\cite{Drummond1980}. These couplings can alter the speed of photon and  gravitational waves   depending  on the spacetime curvature. Thus the spacetime geometry is considered as a special kind of material medium varied the propagation speed. These effects can be used to determine  time delay and structural properties of gravitational waves relative to a vacuum where there is no non-minimal coupling. Furthermore  such $RF^2$ type couplings can be obtained from Kaluza-Klein reduction of a five-dimensional theory \cite{Buchdahl1979,Muller-Hoissen1988,Dereli19901}.
Also, these couplings  can describe the production of primordial magnetic fields, although they do not reach levels close enough to today's values \cite{Turner1988,Lambiase2004}. However it is  important to note that  more generally $R^\alpha F^2$  couplings may   generate  larger   primordial magnetic fields to explain current  observations \cite{Mazzitelli1995,Campanelli2008,Lambiase2008,Bamba2007,Bamba20082,Bamba20081}.
On the other hand, the non-minimally coupled gravity models with electromagnetism and gravity are  very intriguing, since they have solutions to the cosmological and astrophysical problems of gravity. Some properties and spherically symmetric  solutions of these models were given in \cite{Dereli20111,Dereli20113,Sert12Plus,Sert13MPLA,Dereli20112,Sert16regular}. Later  some compact star solutions \cite{Sert2017,Sert2018,Sert20182}  and cosmological solutions  providing important clues in the study of inflation
and the early universe driven by electromagnetic fields \cite{AADS,SertAdak2019} were obtained.
We note that the Lagrangian of this model  with a constraint leads to the second-order derivatives both in the electromagnetic  and gravitational field equations.

In this study we construct traversable wormhole solutions  satisfying the null energy condition   in the non-minimally coupled models.
We give   the solutions  which   have electrically  and/or magnetically charged  and they  can be transformed into each other by the duality transformation of the model. For the asymptotically flat case, we obtain the non-minimal models in $R^{\alpha} F^2$ form satisfying these solutions with  a constant $\alpha$.

\section{Field Equations of the Non-minimally Coupled Gravity} \label{model}
In the presence of matter,  Lagrangian of the model consisting of the non-minimal coupling between gravity and electromagnetism   can be written as follows  \cite{Dereli20111,Dereli20113,Sert12Plus,Sert13MPLA,Dereli20112,Sert16regular,Sert2017,Sert2018}
\begin{equation}\label{action1}
L  =   \frac{1}{2\kappa^2} R*1 -Y(R) F\wedge *F  + L_{mat} + \lambda_a\wedge T^a  \;.
\end{equation}
Here $R$ is the curvature scalar,   $*1 $ is the volume 4-form,  $Y(R)$ is an arbitrary real  function of $R$,     $F=dA$ is  the Maxwell tensor with   potential 1-form $A$,  $L_{mat}$  is the matter Lagrangian of the isotropic fluid, $T^a$ torsion tensor 2-form and $\lambda_a $ is a Lagrange multiplier which leads to a Riemannian geometry without torsion. The isotropic fluid matter assumption significantly simplifies  the results, since it is compatible with the electromagnetic energy-momentum tensor.
By taking variation of this Lagrangian,  the gravitational field equation is found as
 \begin{align}
\label{gfe2}
- \frac{1}{2 \kappa^2}  R^{bc}
\wedge *e_{abc}  =   Y  (F_a \wedge *F - F \wedge \iota_a *F)   -\frac{k}{\kappa^2}*R_a 
+   (\rho +p )u_a*u + p*e_a \ 
 \end{align}
 together with the constraint  
 \begin{eqnarray}\label{cond0}
Y_R F_{mn}F^{mn} = -\frac{ k}{ \kappa^2} \ . 
\end{eqnarray}
Here  $Y_R = \frac{dY}{dR}$,  $p$ is the pressure  and $\rho$ is the energy density of the matter, $u=u_a e^a$ and  $u_a$ is the time-like vector field  for an observer co-moving with the perfect fluid and $k$ is a real  constant which determines the strength of the coupling between gravity with electromagnetism.

It  is important to note that  the gravitational field equation (\ref{gfe2}) of the model does not involve derivatives higher than second-order by  the constraint (\ref{cond0}) and therefore this field equation does not have any  ghost
degrees of freedom.
A second way to obtain the second-order field equation for this model is to add this constraint (\ref{cond0}) to the action of the model (\ref{action1})   by a Lagrange multiplier
and the new action gives  the field equation (\ref{gfe2}) after the variation.

We remember  that the field equations of the non-minimal model
have the  duality transformation which is  $(F,*F)\rightarrow (*YF, -YF) $ and $(Y\rightarrow \frac{1}{Y})$ and it is the modified version of the duality transformation of Maxwell  equations that is  $(F,*F)\rightarrow (*F, -F) $.
It is remarkable to note that while  this duality transformation makes the field equations  (\ref{gfe3}) and (\ref{Maxwell1}) invariant,  it causes  the modified Maxwell Lagrangian  to change sign, $YF\wedge *F \rightarrow - YF\wedge *F $, as in the standard Maxwell Lagrangian, $F\wedge *F \rightarrow - F\wedge *F $. Since the field equations are invariant, all possible solutions to these equations have this duality property.
Therefore it gives  a map
between the electromagnetic tensor $F$ with $*Y (R)F$ under which all the  solutions are mapped into new  solutions. 
The modifications of the Maxwell tensor by the non-minimal function $Y(R)$  can be expressed  with the   constitutive tensor $G=YF$  
\cite{Dereli2007,Dereli20072,Dereli20113}.  Then these modified Maxwell equations describe  a specific medium with a polarization and magnetization determined by the non-minimal function $Y(R)$.

The gravitational  field equation  (\ref{gfe2}) describing our model can be written  in an equivalent way as 
\begin{eqnarray}\label{gfe3}
    G_a=\kappa^2\tau_a^{eff}
\end{eqnarray}
where  
$G_a = - \frac{1}{2 \kappa^2}  R^{bc}   \wedge *e_{abc} = *R_a-\frac{1}{2}R*e_a$ is the Einstein tensor
and $ \tau_a ^{eff}$ is   the effective energy-momentum tensor given by  
\begin{eqnarray}\label{tauabeff}
   \tau_a ^{eff}&=&  Y  (F_a \wedge *F - F \wedge \iota_a *F)   -\frac{k}{\kappa^2}*R_a 
+   (\rho +p )u_a*u + p*e_a \\
&=&  \kappa^2\tau_a^{NM} +\kappa^2{\tau_a}^{matt}  = \tau_{ab}^{eff}*e^b
\end{eqnarray}
as the sum  of the non-minimal energy-momentum tensor $\tau_a^{NM}= Y  (F_a \wedge *F - F \wedge \iota_a *F)   -\frac{k}{\kappa^2}*R_a $ and matter energy-momentum tensor $\tau_a^{matt} =  (\rho +p )u_a*u + p*e_a $. The non-minimal   energy-momentum tensor involves the electromagnetic  part scaled by the function $Y(R)$ and the interaction part  $\frac{k}{\kappa^2}*R_a$  that is purely geometric.
Here we  define that $\tau_{00}^{eff}= \rho^{eff}$ is the effective energy density,  $\tau_{11}^{eff}= P_{r}^{eff}$ is the effective radial pressure,  ${\tau_{22}}^{eff}= \tau_{33}^{eff}= P_{t}^{eff}$ is the effective tangential  pressure, while 
${\tau_{00}}^{matt}=\rho$ and ${\tau_{11}}^{matt}={\tau_{22}}^{matt}={\tau_{33}}^{matt}= p$  are the energy density and pressure for matter. 
The isotropic matter assumption 
significantly simplifies the analysis of the energy conditions, as this means
it is of Hawking-Ellis type 1 (or the corresponding Segre class) \cite{Visser2021,Curiel2017,Hawking1974}. Here we could assume the existence of anisotropic rather than isotropic matter. But in this case, we would have added an extra unknown function to this system of differential equations. In order to close this system, it was necessary to add an equation of state that was a relationship between radial  or tangential  pressure and energy density. However,  we were unable to obtain a meaningful wormhole geometry for this case.

Since the covariant exterior derivative of the Einstein tensor is automatically zero, $DG_a=0$, this effective energy-momentum tensor is conservative $ D\tau_a ^{eff}=0$.
There are also the following modified Maxwell field equations 
\begin{eqnarray}
d(*Y F) = 0, \hskip 1.5 cm dF =0 \; .   \label{Maxwell1}
\end{eqnarray} 
Additionally, we can obtain
 \begin{eqnarray} \label{trace}
 \frac{1-k}{\kappa^2} R = (\rho  - 3p) \; 
 \end{eqnarray}
 by taking  the trace of the gravitational field  equation (\ref{gfe2}).

\section{WORMHOLE  SOLUTIONS}
We consider the following spherically symmetric line element  to describe  the traversable wormholes in 3+1 dimensions: 
\begin{eqnarray}\label{metric}
ds^2 & =& -f^2(r)dt^2  + \frac{dr^2}{1- b(r)/r} + r^2d\theta^2 +r^2\sin^2\theta d \phi^2 
\end{eqnarray}
Here the function $f(r)$ is known as the redshift function and it must be finite  or not have an event horizon in the range from the  throat $r_0$ to infinity in order to obtain traversable wormholes. The other metric function $b(r) $  is known as the shape function and it must satisfy the condition $(b-b'r)/b^2 >0 $  
for traversability of wormholes \cite{MorrisThorne1988,Hochberg1997}.
This condition corresponds to  a flaring out surface, that is  getting wider smoothly  in diameter from  throat of  wormhole.
  The condition becomes $b'(r_0) < 1$  for $r=r_0$ and $b(r)<r$ for $r>r_0$. Also a wormhole metric   must be asymptotically flat or asymptotically dS/AdS. 
In accordance with the  spherical symmetry of the geometry, we will consider  the following  Maxwell tensor 2-form which  has electric and magnetic field component in  radial direction   
\begin{eqnarray}
\label{Maxwellanzats}
 F &= & E(r) e^1\wedge e^0 +B(r)e^2\wedge e^3 .
\end{eqnarray}
The electric field $E(r)$ or magnetic field $B(r)$ can be turned off  depending on the source  by setting $E=0$ or $B=0$ in the resulting differential equations, respectively. 
For this ansatz, the constitutive tensor becomes
$G=YF = YE e^1\wedge e^0 + YBe^2\wedge e^3  $, where we can define  the displacement field $D=YE$ and the  magnetic field  $H=YB$ in the material medium. Then the modified Maxwell equations (\ref{Maxwell1}) lead to the following  equations.
\begin{eqnarray}\label{Maxwell2}
 D(r)=   Y(r)E(r) = \frac{q_e}{r^2}\;, \hskip 1.5 cm B(r)=\frac{q_m}{r^2}
\end{eqnarray}
Here $q_e$ and $q_m$  are integration constants corresponding to the electric and magnetic monopole charge of the source.    These ansatz give us the following differential equations for the gravitational field equation 
(\ref{gfe2})
 \begin{eqnarray}
   \frac{b'}{r^2}  &=&\rho^{eff} 
   \;, \label{d1} 
\\
  (1 - \frac{b}{r} )\frac{2f'}{rf}  - \frac{b}{r^3} &=& P_r^{eff}
   \;,  \label{d2}
\\
  \frac{f''}{f}  ( 1-\frac{b}{r})  + \frac{f'}{rf} ( 1-\frac{b'}{2}  -\frac{b}{2r} ) -\frac{b'}{2r^2}  +\frac{b}{2r^3}   &=&  P_t^{eff}   \;, \label{d3}
\end{eqnarray}
with the constraint
\begin{eqnarray}
    \frac{dY}{dR}(E^2-B^2) = \frac{k}{2\kappa^2}\label{constraint}\;.
\end{eqnarray}
Here we have defined
\begin{eqnarray}
  \rho^{eff} &=& - k\Big[( \frac{f''}{f} + 2\frac{f'}{rf} )( 1-\frac{b}{r})  + \frac{f'}{2f}(-\frac{b'}{r}   + \frac{b}{r^2})\Big] +  \kappa^2 Y( E^2+B^2)     + \kappa^2\rho \;, \\
    P_r^{eff}&=&   k\Big[ ( \frac{f''}{f} ( 1-\frac{b}{r})  + (\frac{f'}{2f}+\frac{1}{r} )(-\frac{b'}{r}  + \frac{b}{r^2}) \Big]
- \kappa^2 Y( E^2+B^2)     +  \kappa^2 p \;,
    \\
     P_t^{eff}  &=&  -k\Big[ - \frac{f'}{rf} ( 1-\frac{b}{r})  + \frac{b'}{2r^2} 
 + \frac{b}{2r^3}   \Big] +\kappa^2 Y( E^2+B^2)     +  \kappa^2 p \;,
\end{eqnarray}
then the Ricci curvature scalar becomes
\begin{eqnarray}\label{Ricci1}
    R = \frac{2f''b}{rf} -\frac{2f''}{f} +\frac{f'b'}{rf} + +\frac{3f'b}{r^2f} - \frac{4f'}{rf}+\frac{2b'}{r^2}\;.
\end{eqnarray}
In order to find solutions to this system of  (\ref{Maxwell2},\ref{d1},\ref{d2},\ref{d3},\ref{constraint})  differential equations, we need to close it by determining the non-minimal function $Y(R)$.
However,  when we examine this system, finding a shape function $b(r)$ that gives appropriate conditions for the wormhole geometry  after determining $Y(R)$ leads to very complicated and tedious calculations.  
Therefore, instead of determining $Y(R)$ function, similar to the reverse-engineering approach in 
\cite{Sebastiani2011,Capozziello2007},
we try to solve this system by choosing the $f(r)$ function appropriately. 
In the  Morris–Thorne paper, \cite{MorrisThorne1988}, tidal effects are actually quantified to avoid stretching of travelers.  Although,  this is not the same as setting the tidal effects to zero, we could not find any solution other than this choice.
  Therefore we consider   the simple redshift function $f(r) = f_0=constant $,   which give regular solutions and  zero tidal force for stationary observers, as taken in the appendix of the inspiring  paper \cite{MorrisThorne1988}.
For $f(r) = f_0=constant $ case,  the above equations turn out to be 
\begin{eqnarray}
 \frac{b'}{r^2}  &=&    \kappa^2 Y( E^2+B^2)     + \kappa^2\rho   = \rho^{eff}  \;, \label{df1}
\\
   - \frac{b}{r^3} &=&  k (-\frac{b'}{r^2}  + \frac{b}{r^3}) - \kappa^2 Y( E^2+B^2)     + \kappa^2 p =   P_r^{eff} 
    \;, \label{df2}
\\
  -\frac{b'}{2r^2}  +\frac{b}{2r^3}     &=&    - k (\frac{b'}{2r^2} 
 + \frac{b}{2r^3}   )+ \kappa^2 Y( E^2+B^2)     +  \kappa^2 p   = P_t^{eff} \;, \label{df3}
\end{eqnarray}

with the constraint equation (\ref{constraint}).  The Ricci scalar (\ref{Ricci1}) simplifies to
$    R = \frac{2b'}{r^2}$. Here we see that $P_t^{eff} =-\frac{1}{2}(\rho^{eff}+  P_r^{eff} )$.

\subsection{Solutions with Electric and Magnetic  Field}

 We firstly look at   a generic solution with electric and magnetic field 
to these equations (\ref{Maxwell2},\ref{constraint},\ref{df1},\ref{df2},\ref{df3}). 
Then we find the following  set of solutions for the unknown functions:
\begin{eqnarray}
b(r) &=&   \frac{C_1}{6}r^3  + \frac{C_2(3k-1)}{2(k+1)} r^{\frac{k+1}{3k-1}}  \;, \label{b1} \\
Y(r) &=& \frac{ C_2(2k-1)r^{\frac{4k}{3k-1}} \pm \sqrt{(2k-1)^2C_2^2r^{\frac{8k}{3k-1}}-16\kappa^4q_e^2q_m^2  }}{4\kappa^2 q_m^2 }\;, \label{Y1}\\
p(r) &=& \frac{(2k-1)C_1}{6\kappa^2}\;,\label{ps1}\\
\rho(r) &=& \frac{C_1-2C_2(k-1) 
r^{\frac{-8k+4}{3k-1}}}{2\kappa^2} \label{rhos1}
\end{eqnarray}
with  the Ricci scalar
\begin{eqnarray}\label{R21}
     R = C_1+C_2 r^{-\frac{4(2k-1)}{3k-1}}\; 
\end{eqnarray}
 where $C_1$ and $C_2$ are integration constants.
 We have also the following  displacement field  and magnetic field from Maxwell equations.
\begin{eqnarray}
    D(r) = YE = \frac{q_e}{ r^2},   \hskip 1 cm B(r)=\frac{q_m}{r^2} \label{DandB}
\end{eqnarray}  
 It is important to emphasize that since this system of differential equations (\ref{df1},\ref{df2},\ref{df3}) is nonlinear, it may have multiple solutions.  The square root with minus and plus sign in $Y(r)$ (\ref{Y1}) comes from  the term 
   $Y(E^2 + B^2) $ by virtue of (\ref{DandB}).
We  can find the inverse function $r(R)$ from (\ref{R21}) as
\begin{eqnarray}
r =   ( \frac{R-C_1}{C_2})^{\frac{3k-1}{4(1-2k)}} 
\end{eqnarray}
and substituting  it into (\ref{Y1}), the non-minimal function becomes 
 \begin{eqnarray}
     Y(R) = \frac{(2k-1)  C_2^{\frac{3k-1}{2k-1}} (R-C_1)^{\frac{k}{(1-2k)}} \pm \sqrt{(2k-1)^2C_2^{\frac{6k-2}{2k-1}}( R-C_1)^{\frac{2k}{(1-2k)}}-16\kappa^4q_e^2q_m^2  }}{4\kappa^2 q_m^2 }\label{YRem}
 \end{eqnarray}
 in terms of Ricci scalar $R$.
 The shape function $b(r)$ (\ref{b1}) 
satisfies 
the condition $b(r_0) = r_0 $  for $C_1=c_1 /r_0^2$ and $C_2= (1-\frac{c_1}{6})\frac{2(k+1)}{3k-1}r_0^{\frac{2k-2}{3k-1}}$. Then the shape function becomes
 \begin{eqnarray}
     b(r) = \frac{c_1}{6r_0^2}r^3 -  \frac{(c_1-6)r^{\frac{k+1}{3k-1}}}{6 r_0^{\frac{-2k+2}{3k-1}}}
 \end{eqnarray}
and $b'(r_0) <1$ condition  leads to the following inequality
\begin{eqnarray}
   b'(r_0) = \frac{c_1(4k-2) +3(k+1)}{9k-3} <1\;.
\end{eqnarray}
We have  the inequality  $\frac{1+k}{3k-1}<1$ which has the solution $\{ k<\frac{1}{3}, k>1\}$ for asymptotically dS/AdS geometries.   Here the  case with $c_1<0$ gives AdS type solutions, while $c_1>0$  gives dS type solutions. 
In the case of normal matter, 
in order to see whether the null energy condition  $\tau_{ab}k^ak^b \geq 0 $  ($\rho + p \geq 0$) and the weak energy condition $\tau_{ab}u^au^b \geq 0 $  ($\rho\geq 0 $ and $\rho + p \geq 0$)  are satisfied,  we  calculate the following functions 
\begin{eqnarray}
    \rho &=& \frac{c_1}{2r_0^2}  + \frac{(c_1-6)(k^2-1)r^{\frac{k+1}{3k-1}}}{3(3k-1)r^3r_0^{\frac{-2k+2}{3k-1}}} \;, \\
    \rho + p &=& \frac{c_1(k+1)}{3r_0^2} +  \frac{(c_1-6)(k^2 - 1)r^{\frac{k+1}{3k-1}}}{3(3k-1)r^3r_0^{\frac{-2k+2}{3k-1}}}\;.
    \end{eqnarray}
  They will monotonically approach to their first term  as $r\rightarrow \infty$ from the initial value
   at $r=r_0$.  
  Then the null energy condition $\rho + p \geq 0$ is satisfied in the  range of the parameters  $\{c_1<0, k<-1\}$.
But we can not find any parameter range that satisfies the weak energy condition, since the asymptotic limit of $\rho$  gives the inequality $c_1/2r_0^2>0$.
Also we check the effective energy density and pressures by using the definitions  (\ref{df1},\ref{df2}, \ref{df3}).
\begin{eqnarray}
    \rho^{eff} &=& \frac{c_1}{2r_0^2} -  \frac{(c_1-6)(k+1)r^{\frac{k+1}{3k-1}}}{6(3k-1)r^3r_0^{\frac{-2k+2}{3k-1}}}, \label{rhoeff}\\
    P_r^{eff}  &=&   -\frac{c_1}{6r_0^2}- \frac{(c_1-6)r^{\frac{k+1}{3k-1}}}{6r^3 r_0^{\frac{-2k+2}{3k-1}}},\\
        P_t^{eff}  &=&   -\frac{c_1}{6r_0^2}   + \frac{(c_1-6)(1-k)r^{\frac{k+1}{3k-1}}}{2(3k-1)r^3 r_0^{\frac{-2k+2}{3k-1}} }\,.
\end{eqnarray}
Since $\frac{k+1}{3k-1}<1$,  the second terms in the functions goes to zero monotonically  at the limit $r\rightarrow \infty$. Hence in order for the energy density  to be positive,     $c_1>0$  must be and  $  \rho^{eff}$ approaches to the limit $\frac{c_1}{2r_0^2}$ from its maximum value at $r_0$.
When we calculate  the energy conditions   \begin{eqnarray}
      \rho^{eff}  + P_r^{eff} = 
     \frac{c_1}{3r_0^2} + \frac{(c_1-6)(k+1)r^{\frac{k+1}{3k-1}}}{3(3k-1)r^3r_0^{\frac{-2k+2}{3k-1}}} \;,\label{rhopreff}\\
      \rho^{eff}  + P_t^{eff} =   \frac{c_1}{3r_0^2} -  \frac{(c_1-6)r^{\frac{k+1}{3k-1}}}{3(3k-1)r^3r_0^{\frac{-2k+2}{3k-1}}} \;,\label{rhopteff}
      \\
         \rho^{eff}  + P_r^{eff}+ 2P_t^{eff}  =0\;.
\end{eqnarray}
  The parameters range $ \{ k<1/3, k\geq 1, c_1>\frac{3(k-1)}{2k-1}\}$ satisfies  the null and weak energy conditions. 
Thus we see that while the matter energy-momentum tensor violates the weak energy condition, the effective tensor   satisfies for some parameter values. However the null energy condition is satisfied by the both tensors  
in the  range of the parameters  $\{c_1<0, k<-1\}$.

If we look at solutions with only magnetic  field $B(r)=q_m/r^2$  by setting $E=0$ or $q_e=0$  in the non-minimal model with the function (\ref{YRem}), then we find   that  
\begin{eqnarray}
Y(R) = \frac{(2k-1)  C_2^{\frac{3k-1}{2k-1}} (R-C_1)^{\frac{k}{(1-2k)}} }{2\kappa^2 q_m^2 }\label{YRm}\;.
\end{eqnarray}

For the case  with only electrically charged solutions, we have the displacement field $D(r) = Y(r)E(r) = q_e/r^2$.
We see that   $q_m=0$ makes the non-minimal function  (\ref{YRem}) singular for some cases and indeterminate for others, depending on the sign of the first term in the numerator.
Instead, when we set $ B(r)=0$ in the Maxwell ansatz  (\ref{Maxwellanzats}), which corresponds to setting $ B(r)=0$   in the (\ref{Maxwell2},\ref{constraint},\ref{df1},\ref{df2},\ref{df3}) equations,
this non-minimal function is found as the solution of this system of  equations as follows:
\begin{eqnarray}
Y(R) = \frac{2\kappa^2 q_e^2 }{(2k-1)  C_2^{\frac{3k-1}{2k-1}} (R-C_1)^{\frac{k}{(1-2k)}} }\label{YRe}\;
\end{eqnarray}
which is  inverse of the previous  one (\ref{YRm}) with $q_m=q_e$. 
As a second method, this function can also be found by performing the modified duality transformation  $(E,B)\rightarrow (YB, -YE) $ and $(Y\rightarrow \frac{1}{Y})$ in the first magnetic solution.  Since $E=0$ in the first solution, the mapping   ($B\rightarrow -YE) $ and $(Y\rightarrow \frac{1}{Y})$ transform the previous magnetic field $B=q_m/r^2$ to the new displacement field  $D=YE =q_e/r^2$ with $q_m\rightarrow -q_e$, while transforming the previous non-minimal function $Y(R)$ to the current inverse function $1/Y(R)$.
 It is remarkable to note that while  this duality transformation   $(F,*F)\rightarrow (*YF, -YF) $ and $(Y\rightarrow \frac{1}{Y})$ makes the field equations  (\ref{gfe3}) and (\ref{Maxwell1}) invariant,  it causes  the modified Maxwell Lagrangian  to change sign, \begin{eqnarray}
    YF\wedge *F \rightarrow \frac{1}{Y}(*YF)\wedge (-YF) = - YF\wedge *F 
\end{eqnarray}
as in the standard Maxwell Lagrangian, $F\wedge *F \rightarrow - F\wedge *F $. Since the field equations are invariant, all possible solutions to these equations have this duality property.
Therefore it gives  a map
between the electromagnetic tensor $F$ with $*Y (R)F$ under which all the  solutions are mapped into new  solutions. 
Also the modifications of the Maxwell tensor by the non-minimal function $Y(R)$  can be expressed  with the   constitutive tensor $G=YF$.   The polarization and  magnetization effects of this kind of a specific medium are  described by   the non-minimal  $Y(R)$ function and  their intensity is determined by the dependence of the $Y(R)$ function on $R$.

\subsection{Asymptotically Flat Solutions} 
For asymptotically flat geometries, we have to set  $C_1=0$ or $c_1 =0$ in the solutions. 
Then the first flare-out condition $b(r_0)=r_0$ determines the other integration constant $C_2$ as
\begin{eqnarray}\label{C2}
    C_2= \frac{2(k+1)r_0^{\frac{2(k-1)}{3k-1}}}{3k-1}\;.
\end{eqnarray}
Also the second flare out condition gives us the inequality $b'(r_0) =  {\frac{k+1}{3k-1}}< 1 $ at the throat,
which has the solution $ \{ k<1/3, \ k> 1\}$.
By substituting these integration constants $C_1$ and $C_2$ to the previous asymptotically AdS/dS solutions   we  obtain
\begin{eqnarray}
   b(r) &=& r_0 (\frac{r}{r_0})^{\frac{k+1}{3k-1}} \;, \label{b2}\\
   R (r) &=& \frac{2(k+1) r_0^{\frac{2(k-1)}{3k-1}} }{3k-1} r^{-\frac{4(2k-1)}{3k-1}}\label{R2} \;, \\
Y(r) &=& \frac{\kappa^2 q_e^2 (3k-1) r_0^{-\frac{2(k-1)}{3k-1}} }{(k+1) (2k-1) }r^{-\frac{4k}{3k-1}} \;,\label{Y2}
 \\
D(r) &=& Y(r)E(r) = \frac{q_e}{r^2}  \;,\label{E2} \\
p(r) &=& 0\label{p2}\\
\rho(r) &=& \frac{2(1-k^2) r_0^{\frac{2(k-1)}{3k-1}}}{(3k-1)\kappa^2} r^{-\frac{4(2k-1)}{3k-1}}\;.\label{rho2}
\end{eqnarray}

A similar wormhole geometry given by the metric function (\ref{b2})  is studied also in the context of general relativistic phantom energy \cite{Francisco2013,Rahaman2009}, $f(R)$ gravity  \cite{Capozziello2021} and Rastall gravity \cite{Moradpour2017}. 
It is worth emphasizing that the traversable wormhole geometry can be constructed by matching this interior solution with the exterior Reissner-Nordström  solution at  a radius $r=a>r_0$. Then  this constant redshift function   can be set to be
$f_0= 1-\frac{2M}{a} + \frac{q_e^2}{a^2}$  on the boundary and in the interior region similar to the method in \cite{Rahaman2009}. 
Also, the  wormholes  can be constructed  by using the cut-and-paste procedure that matches this internal and the expected external geometry via the junction condition.   The stability analysis of such thin-shell traversable wormholes with this cut-and-paste procedure   has been examined in \cite{Lobo2005,Lobo20052,Eiroa2007,Eiroa2008,Mazharimousavi2010,Yue2011,Ishak2002,Falco2020}.

In normal matter case,
due to $p=0$ in the solution, the energy density $\rho(r)$ should be positive in order  to satisfy these  null and weak energy conditions.
We obtain the inequality 
$ \frac{1-k^2}{3k-1} \geq 0$ from the condition on the energy density (\ref{rho2}) and it has the solution interval  $ \{ k\leq -1, \ \frac{1}{3} < k\leq 1\}$ for the $k $ parameter. 
Also  the function $Y(r) $   is  indefinite for $k=-1$  value. Then   the intersection  of these two intervals obtaining from the flare out conditions and the   energy conditions is  $k<-1$. Thus we see that the solutions in the range $k<-1$  allow wormholes involving normal matter, while  the solutions in the range $k>1$ lead to wormholes involving exotic matter.

On the other hand,
in  modified gravity models the generalized null energy condition  may be violated
in order for these models to have wormhole geometries \cite{Hochberg1997,Lobo2009,Garcia2010,Capozziello2012,Francisco2013,Harko2013,Capozziello2014,Capozziello2015}.  This model also violates  the condition in the asymptotically flat case   for the effective energy-momentum tensor (\ref{tauabeff}), since it is not possible to satisfy these two conditions in the same range by taking $c_1=0$   in the  equations (\ref{rhoeff},\ref{rhopreff},\ref{rhopteff}).  Thus the modified terms by the non-minimal couplings induce  a surrogate to the exotic matter that  violates the energy  condition.
 In this case the non-minimal function $Y(R)$ becomes
\begin{eqnarray}\label{Ye}
    Y(R)= R_0R^\alpha
\end{eqnarray}
where $\alpha =\frac{k}{2k-1}$ and $R_0$ is
\begin{eqnarray}\label{R0}
   R_0= \frac{q_e^2\kappa^2}{2k-1}\left( \frac{3k-1}{k+1}\right)^{\frac{3k-1}{2k-1}}r_0^{\frac{-2(k-1)}{2k-1}}2^{\frac{k}{1-2k}}  \;.
\end{eqnarray}
Then the corresponding  gravitational model to the electrically charged solutions become
\begin{equation}\label{actionson}
L  =   \frac{1}{2\kappa^2} R*1 -R_0R^\alpha F\wedge *F  + L_{mat} + \lambda_a\wedge T^a  \;.
\end{equation}

Here we see that the $\alpha$ parameter, which is the power of the non-minimal term $R^\alpha F\wedge *F$,  takes values $\frac{1}{3} <\alpha <\frac{1}{2}$ for normal matter ($k<-1$), while it takes values from  $\frac{1}{2} <\alpha <1$  for exotic matter ($k>1$).  Also, the model transforms into $R^{-\alpha} F\wedge *F$  in the presence of  magnetic field instead of electric field. 
It is important to remark that the  $R F\wedge *F$-type coupling can describe the production of primordial magnetic fields, although they do not reach levels close to the measured values \cite{Turner1988,Lambiase2004}. However more general $R^\alpha F\wedge *F$-type couplings in (\ref{actionson})   may   generate  larger   primordial magnetic fields \cite{Mazzitelli1995,Campanelli2008,Lambiase2008,Bamba20082}.
Even if $R$ tends to zero at late times, the energy density of the magnetic field becomes increasingly larger. Therefore, such non-minimal $R^{-\alpha} F^2$ type models can increase the strength of  magnetic fields of the universe to very large values \cite{Bamba20081}. General expressions  of the generated magnetic field spectrum and the spectral index were derived in terms of the non-minimal coupling function $Y(R)$ then found some conditions  for the
generation of  sufficiently large magnetic fields 
in  $Y(R)F^2$ models \cite{Bamba2007}.

Another important point of these solutions, as well as  traversability, is stability. To ensure their stability, it is important to  calculate  the adiabatic sound speed in addition to energy conditions and flare-out conditions. 
The condition that the speed of sound is zero at the throat $c_s^2= \frac{dp}{d\rho}(r_0) =0$, can be used to guarantee how perturbations affect solutions and   to obtain stable wormhole geometries  \cite{Capozziello2021}, see also \cite{Luongo2018,Hassani2020,Luongo2014} for applications of this condition in fluid dynamics and cosmology.   From (\ref{ps1},\ref{rhos1}), and (\ref{p2},\ref{rho2}) we see that our solutions automatically satisfy this condition.

\section{Conclusion}
 
We have investigated  wormhole solutions in the non-minimally coupled gravity model with electromagnetism. We have first  constructed asymptotically dS/AdS  traversable wormhole geometries  involving electric and/or magnetic charges.  
In this case, 
while the matter energy-momentum  tensor cannot satisfy the  weak energy condition, the effective  energy-momentum  tensor  satisfies the condition. However the null energy condition is satisfied in both cases for some parameter values.
We would like to emphasize that the obtained solutions comply with the condition that the speed of sound is zero, which makes the isotropic fluid perturbations negligibly small  at the throat and gives stable solutions.
We have also uncovered some asymptotically flat traversable wormhole solutions
 with a realistic pressure-less dust matter for some values of  the parameter $k$ in the range $k<-1$.
 However, for this case, 
 while the matter energy-momentum  tensor satisfies the null and weak energy conditions for some parameter range, the effective  energy-momentum  tensor cannot satisfy the conditions for any range.

We have found solutions  which   have electrically  and/or magnetically charged  and they  can be transformed into each other by the duality transformation of the model. In this model the non-minimal $Y(R)$ function describes the strength of the coupling between electromagnetic and gravitational fields.  The gravitational fields  act as a special kind of material medium for the electromagnetic fields. The polarization and magnetization properties of this medium are encoded by the function $Y(R)$. Thus, this function may vary depending on the intensity of the gravitational field in different phenomena such as wormhole, black hole, neutron star and quasar. 
Moreover, this function may respond differently to the dielectric and magnetic susceptibility of the medium.  Therefore, determining the non-minimal function   describing the gravitational and electromagnetic phenomenons gives us important clues about the properties of this medium.

\appendix

\section*{Appendix}

\section{Index-form of the
action and the field equations.}
In terms of a local inertial coordinate system $\{x^\mu\}$, the  Lagrangian of the model (\ref{action1}) can be written as
\begin{equation}
L  =   \frac{1}{2\kappa^2} R -  \frac{1}{2}Y(R) F_{mn} F^{mn}  + L_{mat}  \;
\end{equation}
by using the Maxwell invariant $F\wedge *F = \frac{1}{2}F_{mn}F^{mn}$. Then the gravitational field 
 equation (\ref{gfe2}) is given  an equivalent index-form 
\begin{eqnarray}
R_{\mu \nu} -\frac{1}{2}g_{\mu \nu} R =\kappa^2\tau_{\mu \nu},
\end{eqnarray}
which  can be obtained   by a variation procedure with respect to the metric, where the  energy-momentum tensor is
\begin{eqnarray}
\tau_{\mu \nu } =&& 2Y(g^{mn} F_{\mu n} F_{\nu m} -\frac{1}{4} g_{\mu \nu } F_{mn} F^{mn}) + Y_R F_{mn} F^{mn} R_{\mu \nu}    + (\rho+p)u_\mu u_\nu + p g_{\mu \nu } 
\end{eqnarray}
and the Maxwell equations are
\begin{eqnarray}
  \partial_\mu [ \sqrt{-g} \tilde{F}^{\mu \nu}] =0, \quad   \partial_\mu [ \sqrt{-g} Y F^{\mu \nu}] =0  .
\end{eqnarray}
Here $\tilde{F}^{\mu \nu}$ is dual of $F^{\mu \nu}$ and we have the constraint  equation 
 \begin{eqnarray}
\frac{dY(R(x^\mu))}{dR(x^\mu)} F_{mn} F^{mn} = -\frac{ k}{ \kappa^2} \ 
\end{eqnarray}
which is equivalent of (\ref{cond0}).

\end{document}